\documentclass{article}
\usepackage{amsmath}
\usepackage[dvips]{graphics}
\usepackage{latexsym}
\usepackage{subfigure}
\DeclareFontFamily{OT1}{rsfs}{}
\DeclareFontShape{OT1}{rsfs}{m}{n}{ <-7> rsfs5 <7-10> rsfs7 <10->
; ; ; ; ; ; ; ; ; ; rsfs10}{}
\DeclareMathAlphabet{\mycal}{OT1}{rsfs}{m}{n}

\newcommand{\bea}{\begin{eqnarray*}}
\newcommand{\eea}{\end{eqnarray*}}
\newcommand{\bean}{\begin{eqnarray}}
\newcommand{\eean}{\end{eqnarray}}
\newcommand{\eqs}[1]{Eqs. (\ref{#1})}
\newcommand{\eq}[1]{Eq. (\ref{#1})}
\newcommand{\meq}[1]{(\ref{#1})}

\newcommand{\ppn}[2]{\frac{\partial #1}{\partial #2}}

\newcommand{\grad}{\nabla}

\newcommand{\eqn}{&=&}
\newcommand{\non}{\nonumber \\}
\newcommand{\oh}{\frac{1}{2}}

\newcommand{\sgt}{\sqrt{h}}
\newcommand{\hsp}{\hspace{0.1mm}}

\newcommand{\rt}{R^{(3)}}

\title{General proof of the entropy principle for self-gravitating fluid  in static spacetimes}
\author{Xiongjun Fang\thanks{Email: damiao\_ 2008@mail.bnu.edu.cn }, Sijie Gao\thanks{Corresponding author. Email: sijie@bnu.edu.cn}\\
Department of Physics, Beijing Normal University,\\
Beijing 100875, China}
\begin{document}
\maketitle

\begin{abstract}
We show that for any perfect fluid in a static spacetime, if the Einstein constraint equation is satisfied and the temperature of the fluid obeys Tolman's law, then the other components of Einstein's equation are implied by the assumption that the total entropy of the fluid achieves an extremum for fixed total particle number and for all variations of metric with certain boundary conditions. Conversely, one can show that the extrema of the total entropy of the fluid are implied by Einstein's equation.  Compared to previous works on this issue, we do not require  spherical symmetry for the spacetime. Our results suggest a general and solid connection between thermodynamics and general relativity.\\

PACS number(s): 04.20.Cv, 04.20.Fy, 04.40.Nr

\end{abstract}

\section{Introduction}
The mathematical analogy between laws of black physics and the ordinary laws of thermodynamics leads to the discovery of black hole thermodynamics \cite{b1}-\cite{hawking75}. In the past four decades, black hole thermodynamics has become an important and fascinating subject in general relativity and other theories of gravity\cite{iyer94}-\cite{gong}. A related but different issue is to the study  of the thermodynamics of ordinary matter in curved spacetime, without the presence of black hole. In contrast to the mystic origin of black hole entropy, local thermodynamic quantities of matter in curved spacetime, like entropy density $s$, energy density $\rho$, local temperature $T$, are well defined. Gravity only affects the distribution of these quantities. There are two apparently independent ways to determine the distribution of matter. From the thermodynamic point of view, the fluid should be configured such that its total entropy attains a maximum value. From the gravitational point of view, the distribution of matter must obey Einstein's field equation. Since entropy plays no role in Einstein's equation, there is no guarantee that the two ways should give rise to the same result. However, an early work by Sorkin, Wald and Zhang \cite{wald81} showed that if the total entropy of a spherical radiation is an extremum and the Einstein constraint equation holds, then  the Tolman-Oppenheimer-Volkoff (TOV) equation of hydrostatic equilibrium can be derived, which was originally derived from Einstein's equation. Recently, Gao \cite{gao} extended SWZ's proof from radiation to a general perfect fluid.  This issue has been further explored in the past year \cite{r1}-\cite{r5}.

The above works  reveal a certain relationship between thermodynamics and gravity. But all these results only apply to spherically symmetric spacetimes. It is unclear whether the entropy principle is consistent with general relativity beyond spherical symmetry. In this paper, we propose and  prove two theorems, showing that, under a few natural conditions, the extrema of total entropy is equivalent to Einstein's equation in any static spacetime. A static spacetime admits a timelike Killing vector field which is hypersurface orthogonal. Our proof only involves general properties of spacetime geometry and thermodynamics for ordinary fluid.

It is worth noting that recently a very comprehensive discussion on the equivalence of  thermodynamic equilibrium and Einstein's equation was provided by Green, Schiffrin and Wald \cite{waldfluid}. Although some of the results in \cite{waldfluid} appear to be similar to ours, there are considerable differences in both assumptions and arguments. For instance, a crucial assumption in \cite{waldfluid} is that the spacetime be asymptotically flat, while our theorems apply to any spacetime region, imposing no global conditions on the spacetime. On the other hand, the static condition in our argument is replaced by the more general stationary condition in \cite{waldfluid}.

\section{Properties of perfect fluid in static spacetimes}

 We consider a general perfect fluid as discussed in \cite{gao}. The entropy density $s$ is taken to be a function of the energy density $\rho$ and particle number density $n$, i.e., $s=s(\rho,n)$. From the first law of thermodynamics, one can derive the integrated form of the Gibbs-Duhem relation,
\bean
s=\frac{1}{T}(\rho+p-\mu n)\,, \label{gb}
\eean
where $p$ and $\mu$ represent the pressure and chemical potential, respectively. All the quantities are measured by static observers with four-velocity $u^a$. These observers are orthogonal to the hypersurface $\Sigma$. Therefore, the induced metric on $\Sigma$ is given by
\bean
h_{ab}=g_{ab}+u_au_b\,.
\eean

The stress-energy tensor $T_{ab}$ for perfect fluid takes the form
\bean
T_{ab}=\rho u_au_b+p h_{ab}\,.   \label{sperfect}
\eean

We shall assume that Tolman's law holds, which states that the local temperature $T$ of the fluid satisfies
\bean
T\chi=T_0\,,  \label{tchi}
\eean
where $\chi$ is the redshift factor for static observers and $T_0$ is a constant. This law establishes the relationship between the fluid temperature to the metric, which was also used in \cite{waldfluid}. Now we show that another similar relation for chemical potential $\mu$ is implied by the Tolman law.

It is straightforward to show, from the conservation law $\grad_aT^{ab}=0$ and stationary conditions, that
\bean
\grad_ap=-(\rho+p)A_a\,,
\eean
where $A^a$ is the four-acceleration of the observer.
Since
\bean
u^a=\frac{\xi^a}{\chi}\,,
\eean
where $\xi^a$ is the Killing vector and $\chi$ is the redshift factor,
one can show that
\bean
A_a=\grad_a\chi/\chi\,.
\eean
and thus
\bean
\grad_ap=-(\rho+p)\grad_a\chi/\chi\,. \label{gp1}
\eean

On the other hand, the local first law can also be expressed in the form \cite{gao}
\bean
dp=sdT+nd\mu   \label{gp2}
\eean
Comparing \eqs{gp1} and \meq{gp2}, and using \eqs{tchi} and \meq{gb}, we find
\bean
\frac{\grad_a \mu}{\mu}=-\frac{\grad_a \chi}{\chi}
\eean
which leads to
\bean
\mu\chi =const.
\eean
Thus, from the Tolman's law and local thermodynamic laws for perfect fluid, we derive the constancy of the redshifited chemical potential. This also yields
\bean
\frac{\mu}{T}=const.  \label{mut}
\eean
This relation will be used later.

\section{Two theorems}
In this section, we present two theorems on the relationship between the extrema of  total entropy of fluid and static solutions to Einstein's equation.
\newtheorem{theo}{Theorem}
\begin{theo}
Consider a perfect fluid in a static spacetime $(M,g_{ab})$ and $\Sigma$ is a three dimensional hypersurface denoting a moment of the static observers. Let $C$ be a region on $\Sigma$ with boundary $\bar C$. Assume that the temperature of the fluid obeys  Tolman's law and the Einstein constraint equation is satisfied in $C$. Then the other components of Einstein's equation are implied by the extrema  of the total fluid entropy for fixed particle number and for all variations where $h_{ab}$ and its first derivatives are fixed on $\bar C$.
\end{theo}

{\em Proof.}
The total entropy $S$ is an integral of the entropy density $s$ over the region $C$ on $\Sigma$
\bean
S=\int_C \sgt s(\rho,n)\,,
\eean
where $h$ is the determinant of $h_{ab}$ in any coordinates of $\Sigma$.
Without loss of generality, we can fix the coordinates on $\Sigma$ for all variations. Thus, the variation of total entropy is written in the form
\bean
\delta S=\int_C s\delta \sgt +\sgt\delta s\,.
\eean
Applying  the local first law of thermodynamics,
\bean
Tds=d\rho-\mu dn\,,
\eean
we find
\bean
\delta S\eqn \int_C s\delta \sgt +\sgt\left(\ppn{s}{\rho}\delta\rho+\ppn{s}{n}\delta n \right)\non
\eqn \int_C s\delta \sgt +\sgt\left(\frac{1}{T}\delta\rho-\frac{\mu}{T}\delta n \right) \label{dels}\,.
\eean
The total number of particle $N$ is the integral
\bean
N=\int_C \sgt\ n\,,
\eean
which yields the variation
\bean
\delta N=\int_C \sgt \delta n+n\delta\sgt\,.
\eean
Therefore, the constraint $\delta N=0$ is equivalent to
\bean
\int_C\sgt\delta n=-\int_Cn\delta\sgt\,.
\eean
With this constraint as well as \eq{mut},  \eq{dels} can be written as
\bean
\delta S\eqn \int_C\left(s+\frac{n\mu}{T}\right)\delta\sgt+\sgt\frac{1}{T}\delta\rho\non
\eqn \int_C\frac{\rho+p}{T}\delta\sgt+\sgt\frac{1}{T}\delta\rho\,.
\eean
Using  \cite{waldbook}
\bean
\delta\sgt=\oh\sgt h^{ab}\delta h_{ab}\,,
\eean
we obtain
\bean
\delta S=\int_C \delta L\,,
\eean
where
\bean
\delta L
\eqn \oh \frac{\rho+p}{T}\sgt h^{ab}\delta h_{ab}+\sgt \frac{1}{T}\delta\rho  \label{dltr}\,.
\eean

Our purpose is to derive  $\delta L=0$ from Einstein's equation. First note that the extrinsic curvature of $\Sigma$ is defined by
\bean
\hat B_{ab}\equiv h^c_a h^d_b \grad_d u_c\,,
\eean
where $\grad_a$  is the derivative operator associated with $g_{ab}$, satisfying $\grad_a g_{bc}=0$.
It is then straightforward to show
\bean
\hat B_{ab}=\grad_b u_a+A_au_b\,,
\eean

where $A^a$ is the four-acceleration of the observer.
Since
\bean
u^a=\frac{\xi^a}{\chi}\,,
\eean

thus
\bean
\grad_b u_a=\frac{1}{\chi}\grad_b\xi_a-u_a A_b\,,
\eean
which leads to
\bean
\hat B_{(ab)}=0\,,
\eean
where $B_{(ab)}$ is the symmetrization of $B_{ab}$.
The antisymmetrization of $B_{ab}$ also vanishes due to the fact that $u^a$ is hypersurface orthogonal\cite{waldbook}. Consequently, $\hat B_{ab}=0$ and
\bean
\grad_b u_a=-A_au_b \label{uaab}\,.
\eean
This formula will be very helpful in the later calculation.

One can show that the curvature $\rt_{abc}\hsp^d$ of $\Sigma$ is related to the spacetime curvature $R_{abc}\hsp^d$ by
\bean
\rt_{abc}\hsp^d= h_a^f h_b^g h_c^k h^d_jR_{fgk}\hsp^j\,.
\eean
Note that there would be $\hat B_{ab}$ terms on the right-hand side if the spacetime were not static\cite{waldbook}.

It is not difficult to find that
\bean
\rt_{ab}=R_{ab}+R_{aeb}\hsp^lu^eu_l +R_{fb}u^fu_a
+u^ku_b R_{ak}+u_au_b R_{fk}u^f u^k \label{r3ab}\,,
\eean
and
\bean
\rt=R+2R_{ab}u^au^b  \label{r3}\,.
\eean

To calculate  $\delta\rho$, we start with constraint Einstein's equation
\bean
G_{ab}u^au^b=8\pi T_{ab} u^au^b\label{eind}\,,
\eean
where the stress-energy tensor $T_{ab}$ for perfect fluid has been given in \eq{sperfect}.
Thus
\bean
\rho=\frac{1}{8\pi}G_{ab}u^au^b \label{cein}\,.
\eean
 Together with \eqs{r3ab} and \meq{r3}, we obtain \cite{MTW}
\bean
\rho=\frac{1}{16\pi}\rt \label{rr3}\,.
\eean
This tells us that the variation of $\rho$ is actually determined by the geometry of $\Sigma$. Denote the last term in \eq{dltr} by $\delta L_1$, which then gives
\bean
\delta L_1=\frac{1}{16\pi T}\sgt \delta \rt = \sgt\frac{1}{16\pi T}\left( h^{ab}\delta \rt_{ab}+\rt_{ab}\delta h^{ab}\right) \label{ltr}\,.
\eean
Denote the first term on the right-hand side by $\delta L_1'$, i.e.,
\bean
\delta L_1'=\sgt\frac{1}{16\pi T} h^{ab}\delta \rt_{ab}\,.
\eean
The standard calculation yields (see e.g., \cite{waldbook})
\bean
\delta L_1'= \sgt\frac{1}{16\pi T}D^av_a\,,
\eean
where
\bean
v_a=D^b\delta h_{ab}-h^{bc}D_a\delta h_{bc}\,,
\eean
and $D_a$ is the derivative operator on $\Sigma$ associated with $h_{ab}$. To get $\delta h_{ab}$ as a common factor, we perform integration by parts and find
\bean
\delta L_1'=\sgt\frac{1}{16\pi }D^a(v_a/T)-\sgt\frac{1}{16\pi }v_aD^a(1/T)\,.
\eean

According to the assumption of Theorem 1, the metric and its first derivatives are fixed on $\bar C$. So  we may get rid of the boundary term and obtain
\bean
\delta L_1'\eqn-\frac{1}{16\pi }\sgt v_aD^a(1/T) \non
\eqn -\frac{1}{16\pi }\sgt D^b\left(\delta h_{ab}\right)D^a(T^{-1})+\frac{1}{16\pi }\sgt h^{bc}D_a\left(\delta h_{bc}\right)D^a(T^{-1})\non\,.
&&
\eean

Using integration by parts again and dropping the boundary terms, we have
\bean
\delta L_1'
\eqn \frac{1}{16\pi }\sgt D^bD^a(T^{-1})\delta h_{ab}-\frac{1}{16\pi }\sgt h^{ab}D_cD^c(T^{-1})\delta h_{ab} \label{l22}\,.
\eean
Now $\delta L_1'$ is linear in $\delta h_{ab}$, as desired.

Without loss of generality, we take $T_0=1$ in \eq{tchi} and \eq{l22} becomes
\bean
\delta L_1'= \frac{1}{16\pi }\sgt D^bD^a\chi\delta h_{ab}-\frac{1}{16\pi }\sgt h^{ab}D_cD^c\chi\delta h_{ab}\,.
\eean
Note that
\bean
D_a \chi=\chi A_a \label{dac}
\eean
Thus,
\bean
D_bD_a\chi=A_aD_b\chi+\chi D_bA_a=\chi A_aA_b+\chi D_b A_a\,.
\eean
and
\bean
h_{ab}D_cD^c\chi=h_{ab}(\chi A^c A_c+\chi D_c A^c)\label{ac2}
\eean
So
\bean
\delta L_1'=\frac{1}{16\pi T}\sgt M_1^{ab} \delta h_{ab}\,,
\eean
where
\bean
 M_1^{ab}=A^aA^b+ D^b A^a-h^{ab}( A^c A_c+ D_c A^c) \label{l2qf}\,.
\eean

We calculate
\bean
D_c A^c\eqn h^c_fh^e_c\grad_eA^f\non
\eqn \grad_c A^c+u^eu_f\grad_e A^f\non
\eqn \grad_c A^c+u^eu_f\grad_e(u^b\grad_b u^f)\non
\eqn \grad_c A^c+u^e\grad_e(u_f u^b\grad_b u^f)-u^e(\grad_eu_f)(u^b\grad_b u^f)\non
\eqn \grad_c A^c-A^cA_c\,,
\eean
and  \eq{l2qf} can be rewritten as
\bean
 M_1^{ab}=A^aA^b+ D^b A^a-h^{ab}\grad_c A^c \label{l2q}\,.
\eean
Substituting these results into
\eq{dltr} yields
\bean
\delta L
\eqn \oh \frac{\rho+p}{T}\sgt h^{ab}\delta h_{ab}+\sgt\frac{1}{16\pi T} \left( -\rt\hsp^{ab}\delta h_{ab}+M_1^{ab} \delta h_{ab}\right) \non
\eqn \frac{\sgt}{T}\left(\frac{\rho+p}{2}h^{ab} -\frac{1}{16\pi}\rt\hsp^{ab}+\frac{1}{16\pi}M_1^{ab}\right)\delta h_{ab}
 \label{dl3}\,.
\eean
This shows explicitly that $\delta S$ is determined by the variation of $h_{ab}$. Since $\delta S=0$ by the assumption of Theorem 1, we have
\bean
\frac{\rho+p}{2}h^{ab} -\frac{1}{16\pi}\rt\hsp^{ab}+\frac{1}{16\pi}\left(  A^aA^b+ D^b A^a-h^{ab}\grad_c A^c\right)=0
\eean
So
\bean
8\pi p h^{ab}=R^{(3)ab}-A^aA^b- D^b A^a+h^{ab} \grad_c A^c-8\pi \rho h^{ab}
\eean
Substitution of \eqs{r3ab},\meq{r3} and \meq{rr3} yields
\bean
8\pi p h^{ab}\eqn h^{ac}h^{bd}R_{cd}+h^{ac}h^{bd}R_{ced}\hsp^lu^eu_l-A^aA^b- D^b A^a\non
&+&h^{ab} \grad_c A^c-\oh R h^{ab}-R_{cd}u^cu^d h^{ab}\non
\eqn  h^{ac}h^{bd}R_{cd}-\oh Rh^{ab}-P_1^{ab}-P_2^{ab} \label{proe}
\eean
where
\bean
P_1^{ab}\eqn h^{ab}R_{cd}u^cu^d-h^{ab}( \grad_c A^c) \label{p1ab}\\
P_2^{ab}\eqn-h^{ac}h^{bd}R_{cedl}u^eu^l   \label{p2ab}
 +A^aA^b+ D^b A^a\,.
\eean

Now we  show that $P_1^{ab}$ and $P_2^{ab}$ vanish respectively. We first calculate
\bean
\grad_cA^c\eqn\grad_c (u^b\grad_b u^c) \non
\eqn (\grad_c u^b)\grad_b u^c+u^b\grad_c\grad_b u^c \,.
\eean
Note that
\bean
\grad_c\grad_b u^d-\grad_b\grad_c u^d=-R_{cbe}\!^d u^e\,. \label{gcbu}
\eean
Hence,
\bean
\grad_c A^c= (\grad_c u^b)\grad_b u^c+u^b\grad_b\grad_c u^c+R_{be}u^bu^e\,.
\eean
Then
\eq{p1ab} can be  written in the form
\bean
P_1^{ab}=h^{ab}\left[-(\grad_c u^d)\grad_d u^c-u^d\grad_d\grad_c u^c \right]\,.
\eean
Since $u^a=\xi^a/\chi$ where $\xi^a$ is the Killing vector field, we have
\bean
\grad_c u^c=0 \,.
\eean
So
\bean
P_1^{ab}=h^{ab}\left[-(\grad_c u^b)\grad_b u^c\right] \,.
\eean
By \eq{uaab}, we find immediately
\bean
P_1^{ab}=0 \,.
\eean

To deal with the first term on the right-hand side of \eq{p2ab}, we start from
\bean
R_{cedl} u^eu^l
\eqn u^e \grad_c\grad_e u_d-u^e\grad_e\grad_c u_d\non
\eqn \grad_c(u^e\grad_eu_d)-(\grad_c u^e)(\grad_e u_d)-u^e\grad_e\grad_c u_d\non
\eqn \grad_c A_d-u_cA^eu_e A_d+u^e\grad_e(u_c A_d) \non
\eqn \grad_c A_d+u_c u^e\grad_e A_d+A_d A_c \,,
\eean
where we have used \eq{uaab} repeatedly. Hence, $P_2^{ab}$ in \eq{p2ab} can be written in the form
\bean
P_2^{ab}\eqn-h^{ac}h^{bd}\grad_c A_d -A^aA^b+A^aA^b+ D^b A^a \non
\eqn -D^a A^b+D^b A^a \,.
\eean
\eq{dac} implies that $D^a A^b$ is symmetric in $a,b$ and thus
\bean
P_2^{ab}=0 \,.
\eean
Therefore, \eq{proe} just gives the projection of Einstein's equation into $\Sigma$
\bean
h^{ac}h^{bd}R_{cd}-\oh R h^{ab}=8\pi p h^{ab} \label{peis}
\eean
 This completes the proof of Theorem 1.

In the above proof, we used the Einstein constraint equation \meq{rr3} to derive \eq{dl3}. Then by applying $\delta S=0$, we obtained the spatial components of Einstein's equation. It is not difficult to check that the proof is reversible, i.e., From the projected Einstein's equation \meq{peis}, one can show $\delta L=0$ in \eq{dl3}, which makes the total entropy be an extremum. Thus, we arrive at the following theorem:

\begin{theo}
Consider a perfect fluid in a static spacetime $(M,g_{ab})$ and $\Sigma$ is a three dimensional hypersurface denoting a moment of the static observers. Let $C$ be a region on $\Sigma$ with a boundary $\bar C$ and $h_{ab}$ be the induced metric on $\Sigma$. Assume that the temperature of the fluid obeys  Tolman's law and Einstein's equation is satisfied in $C$. Then the fluid is distributed such that its total entropy in $C$ is an extremum for fixed total particle number and for all variations where $h_{ab}$  and its first derivatives are fixed on $\bar C$.
\end{theo}

Note that the Einstein constraint equation usually refers to
\bean
G_{ab}u^b=8\pi T_{ab} u^b\,,  \label{gtu}
\eean
while we only used its time component, i.e., \eq{rr3}, throughout the paper. The remaining part of \eq{gtu} reads
\bean
G_{cb}u^bh^c_a=8\pi T_{cb} u^b h^c_a \,. \label{gpia}
\eean
By \eq{sperfect}, the right-hand side simply vanishes for perfect fluid. With the help of \eqs{gcbu} and \meq{uaab}, one can show that the left-hand side of \eq{gpia} also vanishes. Thus, \eq{gpia} is automatically satisfied in static spacetimes.

\section{Conclusions}
We have rigorously proven the equivalence of the extrema of the entropy and Einstein's equation under a few natural and necessary conditions. The significant improvement from previous works is that no spherical symmetry or any other symmetry is needed on the spacelike hypersurface. Our work suggests a clear connection between Einstein's equation and the thermodynamics of perfect fluid in static spacetimes.

\section*{Acknowledgements}
We thank Wald, Green and Schiffrin for reading our manuscript and offering valuable comments which help us improve the quality of the paper. This research was supported by NSFC Grants No. 11235003, 11375026 and NCET-12-0054.

\end{document}